\documentclass{article}

\usepackage{authblk}
\usepackage[utf8]{inputenc}
\usepackage[textsize=tiny]{todonotes}
\usepackage{mathtools}
\usepackage{paralist}
\usepackage{bbm}
\usepackage{array}
\usepackage{amsthm}
\usepackage{amssymb}
\usepackage{soul}

\usepackage{multirow}
\usepackage{forloop}
\usepackage{algorithm}  % the wrapper for floats
\usepackage{algpseudocode}  % the layout, it includes algoritmicx
\usepackage{hyperref}
\usepackage[inline]{enumitem}
\usepackage{tabto}
\usepackage{subcaption} 

\graphicspath{{figures/}}

\newcommand{\removedtodo}[2][]{}

\begin{document}
	%\maketitle
	
	%\begin{frontmatter}
	%
	%%% Title, authors and addresses
	%

	\title{Blockchains Meet Distributed Hash Tables: Decoupling Validation from State Storage\\ (Extended Abstract)}
	\author[a]{Matteo Bernardini}
	\author[b]{Diego Pennino}
	\author[c]{Maurizio Pizzonia}
	\affil[abc]{Universit\`a degli Studi Roma Tre, Dipartimento di Ingegneria, Sezione Informatica e Automazione\\Via della Vasca Navale 79, 00146, Roma (Italy) }
	\affil[a]{mat.bernardini@stud.uniroma3.it}
	\affil[b]{pennino@ing.uniroma3.it} 
	\affil[c]{pizzonia@ing.uniroma3.it}
	\date{}

	\maketitle

	%
	%\author{Diego Pennino$^\mathrm{a}$}
	%\author{Maurizio Pizzonia$^\mathrm{c}$ \\Corresponding Author: Maurizio Pizzonia}
	%\address{Universit\`a degli Studi Roma Tre, Dipartimento di Ingegneria, Sezione Informatica e Automazione \\Via della Vasca Navale 79, 00146, Roma (Italy) }
	%\address{$^\mathrm{a}$pennino@ing.uniroma3.it, $^\mathrm{c}$pizzonia@ing.uniroma3.it}
	%%Dipartimento di Ingegneria, Universit\'a degli Studi Roma Tre
	%
	\begin{abstract}
	  The first obstacle that regular users encounter when setting up a
	  node for a public blockchain is the time taken for
	  downloading all the data needed for the node to start operating
	  correctly. In fact, this may last from hours to weeks for the major
	  networks.
	  
	  Our contribution is twofold. Firstly, we show a design that enables
	  mining and validation of new blocks keeping only a very small
	  state.
%	  , provided that transactions are equipped with proper proofs
%	  derived from authenticated data structures that are similar to
%	  those already commonly used in blockchains. 
	  Secondly, we show that
	  it is possible to store the state of the blockchain in a
	  distributed hash table obtaining a wide spectrum of trade-offs
	  between storage committed by the nodes and replication factor. Our
	  proposal is independent from the consensus algorithm adopted, and
	  copes well with transactions that involve smart contracts.
	  
	  \medskip 	\noindent\textbf{Keywords.} Blockchain, Distributed Hash Table, Synchronisation Efficiency, Integrity.
	  
	\end{abstract}

	\section{Introduction}\label{sec:intro}

Cryptocurrencies based on public blockchains started with the promise
for the common people to gain some freedom from institutional and
gov\-ern\-ment-reg\-u\-lat\-ed payment instruments.  Currently, this romantic
objective is hardly met. The reasons are many. The adoption of a
proof-of-work consensus algorithm is usually regarded as a prominent
one due to its big requirements in terms of computational power. The
needs are so high that only focused organisations can afford mining
(e.g., mining pools). This is one of the reasons that prompt
researchers and other communities to focus on the study of lighter
consensus algorithms, e.g., based on some form of proof-of-stake. The
effort in that direction is remarkable and it is likely that this
problem will be practically overcome quite soon. 

Even supposing to have a light consensus algorithm, when installing a
node a common user stumbles upon another major difficulty: the time
taken to download the data needed for the software to properly operate
and, up to a certain extent, the amount of space required for that
data. Note that this is true also with today's technology when the user
decides to install a validate-only node, which may be useful, for
example, to have an independent way of injecting transactions into the
network. The time needed for the first synchronisation varies depending
%This needs in terms of time vary depending 
on the bandwidth,
on the I/O speed of the mass storage and on the CPU speed of the node. 
This is currently in the order of 6-24 hours for Ethereum but may last
weeks for Bitcoin. Further, if the node is shut down for a while, it needs a
certain amount of time before becoming fully
operational again.

In this paper, we present a blockchain design with certain
notable properties with respect to the ability to quickly start a
validating node, which might possibly be a miner. In particular,
we show that it is possible to run a blockchain in which validating
nodes are only mandated to keep the last $n$ received blocks.

In our approach, nodes do not store the whole state of the blockchain but only 
the state changed by transactions in these blocks, while still being able to 
validate blocks with the same level of security of regular blockchains.

The data needed to perform validation of transactions are
retrieved from untrusted storage by the creator of the transaction and
conveyed to the nodes along with it. Security is obtained by equipping
candidate transactions with proofs derived from authenticated data structures, similar to those already commonly used in blockchains. Since
validation is decoupled from the storage of the bulk of the blockchain
data, we are free to store these data where it is more convenient. Our
proposal is to store them in a Distributed Hash Table (DHT) realised by
the same nodes that perform validations. In principle, a node may choose how much storage to commit
for storing blockchain data, from nothing to the whole blockchain dataset.
The burden to query the DHT is left to the creator of the transaction.
This allows us to obtain a wide spectrum of
trade-offs between storage used by nodes and replication
factor. 

The rest of this paper is structured as follows. In
Section~\ref{sec:soa} we review the state of the art. In
Section~\ref{sec:background} we briefly introduce basic concepts. In
Section~\ref{sec:solution} we describe our decoupled approach. In
Section~\ref{sec:discussion} we discuss performances and security. In
Section~\ref{sec:conclusions} we draw the conclusions.

	\section{State of the Art}\label{sec:soa} In 2008 the
\emph{blockchain} technology was introduced as basis for the Bitcoin
cryptocurrency~\cite{nakamoto2008bitcoin}. In that context, blockchain
addressed the problem to verify the correct behaviour of untrusted
nodes in a peer-to-peer network, with respect to the execution of
payment transactions. Following the same approach, many other
blockchain-based systems extended the application spectrum and the kind
of supported transactions.
%~\cite{walport2016distributed}. 
One notable
example is \emph{Ethereum}~\cite{wood2014ethereum}, which supports
transactions that can execute general purpose scripts called
\emph{smart contracts}. 

Scalability is one of the prominent topics in research about blockchain.
Most of the work is focused on obtaining scalable consensus algorithms
in terms of number of nodes and throughput, keeping latency low. 
A good survey of this area is provided by~\cite{chauhan2018blockchain}. It
compares various scalability issues with common blockchains and recent
proposals to overcome scalability limits. 
It concludes suggesting that
the sharding technique seems to be the most viable method for scaling
the blockchain. Essentially, this technique divides the peer-to-peer network 
in several smaller networks dividing up the load (see for example~\cite{kokoris2018omniledger}).

We focus on a very specific practical problem. In current
technologies, a validating node is required to store a large amount of
data and this is one of the difficulties that hinder the wide adoption of
the current blockchain technology. A Bitcoin ``full node'' needs to
store the whole transaction history. Currently, the Bitcoin database 
of a full node occupies more that 150 GB. In Ethereum, blocks assert
not only consensus on a valid set of transactions but also on an
explicitly represented state of the system in terms of amount of money 
associated with addresses and content of persistent variables of smart
contracts. For this reason, contrary to Bitcoin nodes, Ethereum nodes
can validate and mine new blocks without storing the whole transaction
history. Currently, for Ethereum the full history is larger that 1 TB
but Ethereum does not mandate to store it for mining or validating blocks. A node storing
only the current state needs about 150 GB of free space (using geth fast sync).
The time taken to download and validate these data greatly varies
depending on bandwidth, cpu speed and mass storage speed. However it may go
from several hours to several weeks.
This problem is particularly relevant if blockchain is adopted in IoT devices~\cite{danzi2018analysis}.
The adoption of the sharding technique should mitigate this problem.
However, for several reasons, sharding requires a complete
re-engineering with respect to the most common blockchains.
Further, the impact depends on the size of the shards, which may be affected by other
considerations, and on how large the state associated with each shard is.

Blockchain is also used as a notary service. When notarised data is
more than a handful of bytes, the document is usually stored by a
different service and just its hash is recorded in the blockchain. One
possible approach to keep the decentralised characteristic of the
system intact is to store these documents in a peer-to-peer network,
like IPFS~\cite{benet2014ipfs}. These solutions rely on \emph{Distributed Hash
Tables} (\emph{DHTs}). A DHT is a key-value pair storing system that
is decentralised and distributed and guarantees that any participating
node can efficiently retrieve the value associated with a given key
using a lookup service (see for example~\cite{maymounkov2002kademlia}). In~\cite{tamassia2005efficient} an
authenticated DHT is proposed. In this paper, we adopt a DHT to store
the state of the blockchain to relieve a validating node from storing the
state associated with the addresses of the blockchain and permit higher
flexibility in storage commitment.

	\section{Background}\label{sec:background}

In this section, we recall basic concepts, terminology and properties about
\emph{authenticated data structures} (\emph{ADS}), \emph{Blockchain} and \emph{Distributed Hash Tables} (\emph{DHT}), limiting the matter to what
is strictly needed to understand the rest of this paper.

\subsection{Authenticated Data Structures (ADS)}\label{sec:back.ADS}

For this paper, an ADS is a container of 
key-value pairs, which are also called \emph{elements}. 
The ADS deterministically
provides a constant-size digest of its content
with the same properties of a cryptographic hash. We call it \emph{root-hash},
denoted by $r$. If the value of any element of the set changes, $r$ changes. 
An ADS provides two operations, the authenticated query of a key $k$
and the authenticated update of a key $k$ with a new value $v'$. A query
returns the value $v$ and a proof of the result with respect to the
current value of $r$. 
The update operation on $k$ changes $v$ associated with
$k$ into a provided $v'$ and changes $r$ in $r'$, as well. An interesting aspect is
that a trusted entity that intends to update $k$ can autonomously
compute  $r'$ starting from the proof of $\left<k,v\right>$ 
obtained by a query.

A typical ADS is a Merkle Hash Tree in which each leaf stores the hash 
of $\left<k,v\right>$ and each internal node stores the hash of the composition of the hashes of its children.
In this case, a proof for $k$ is constructed by considering the path from 
the leaf associated with $k$ to the root. The proof contains the
sequence of the hashes stored in the siblings of each node in that path
labelled with the indication that the path is entering a node from the left 
or right child. The proof check is performed by computing the hashes on the path starting from the leaf and comparing the result with the root-hash.
To update the root-hash, the same computation is performed using the new value when computing the hash of the leaf.

When we have a large set of elements stored in an ADS, but we only
need authentication for a small number of them, known in advance, we can resort to
the \emph{pruning} technique.
Pruning reduces the storage size of the tree, without changing the root-hash,
by removing sections of the tree that are not needed for the expected queries. 
The basic idea is very simple. Whenever 
a subtree has only unneeded leaves, we can remove all the
subtree maintaining only its root with its original hash. Pruning an ADS reduces the required space, 
preserves the root-hash, preserves the capability to produce proofs for 
the needed keys, and keeps security intact.

Further details can be found in~\cite{tamassia2003authenticated, martel2004general}.

\subsection{Blockchain}\label{sec:back.Blockchain}

%We introduce the basic terminology related to blockchains that is
%strictly needed for the rest of the paper.
From our point of view, a
\emph{blockchain} is a data structure that stores a \emph{state} and
its evolution over time. The state is a sort of key-value store where
keys are called \emph{(state) elements}. The concept of state element
is an abstraction that may be regarded as an \emph{address} with a
corresponding balance (following the Ethereum terminology) or as a
\emph{variable} of a contract account with its value. A
\emph{transaction} is an atomic change of a number of involved state
elements. A block is essentially a sequence of transactions. Each
block is hash-chained with the previous one in a \emph{blockchain}. In
our model, each block is associated with a state before and after the
execution of its transactions. Blocks are generated at regular
intervals of length $T$, called \emph{block time}. Each block is identified by 
an increasing number: its \emph{index}. The block with index $i$ is denoted
$b_i$. The \emph{depth} of a block is the number of blocks that were mined after it plus one. The depth of the last mined block is 1. 
To be \emph{valid},
a block  should conform to a number of \emph{consensus rules}, which
may deal with specific semantic constraints (like accounting
constraints or smart contract execution). Even if consensus rules are
fundamental in practice, the rest of the paper is largely independent
from the specific rules enforced by a blockchain. The \emph{consensus
algorithm} is the way nodes reach an agreement about the next block to
be added to the blockchain. The rest of the paper is independent from
the specific algorithm adopted by a blockchain. Certain algorithms may
temporarily produce \emph{forks}, that is more chains are grown at
the same time for a while, then one of them is chosen (usually the longest one) by all nodes
discarding the blocks of the other chains.

\subsection{Distributed Hash Tables (DHT)}\label{sec:back.DHT}

Distributed Hash Tables are distributed data structures supporting  
 put() and get() primitives on key-value pairs. Most DHT implementations can locate 
an object in $O(\log n)$ network operations, where $n$ is the number of nodes 
of the DHT, and provide a fault-tolerant way to access large amounts of data. 
The \emph{keyspace} is the set of all possible keys. Each node stores a subset 
of key-value pairs among the keyspace. We say a node $N$ is \emph{authority} 
for the key $k$ if it stores its data. Each node also gets assigned an identifier from 
the keyspace. The DHT defines a \emph{distance} function between keys. Typically a 
node $N$ is authority for keys close to its identifier according to that distance.

In our study, we do not need the put() operation. The get($ k $) operation returns the 
value associated with a key $k$, performing a lookup in the network, to locate a node that is authority for $k$. For that purpose a suitable 
routing algorithm is used, with each node storing a routing table based on the distance among node identifiers.

	\section{Block Validation and State Storage as Separate Roles}\label{sec:solution}

In our approach, each node has two distinct roles: the
\emph{storage} role and the \emph{validation} role. For the storage
role, each node stores values for a subset of the state elements,
essentially acting as a DHT node. For the validation role, each node
mines new blocks and validate blocks that are broadcasted in the
network. Contrary to the traditional approach, validation does not rely on local storage of the state.
A node may not play the storage role at all.
Further, any node
can create a new transaction and broadcast it so that it can be
included (if valid) in one of the next blocks during mining.

\subsection{Storage Role of a Node}

For this role, each node acts as a node of a DHT. The whole DHT stores
the state of the blockchain (see Section~\ref{sec:back.Blockchain}) and is able to reply to each query for a state element $e$ with
its value $v$.

\begin{figure}
		\centering	
	\begin{minipage}[b]{\linewidth}
		\centering	
		
		\includegraphics[width=0.9\linewidth]{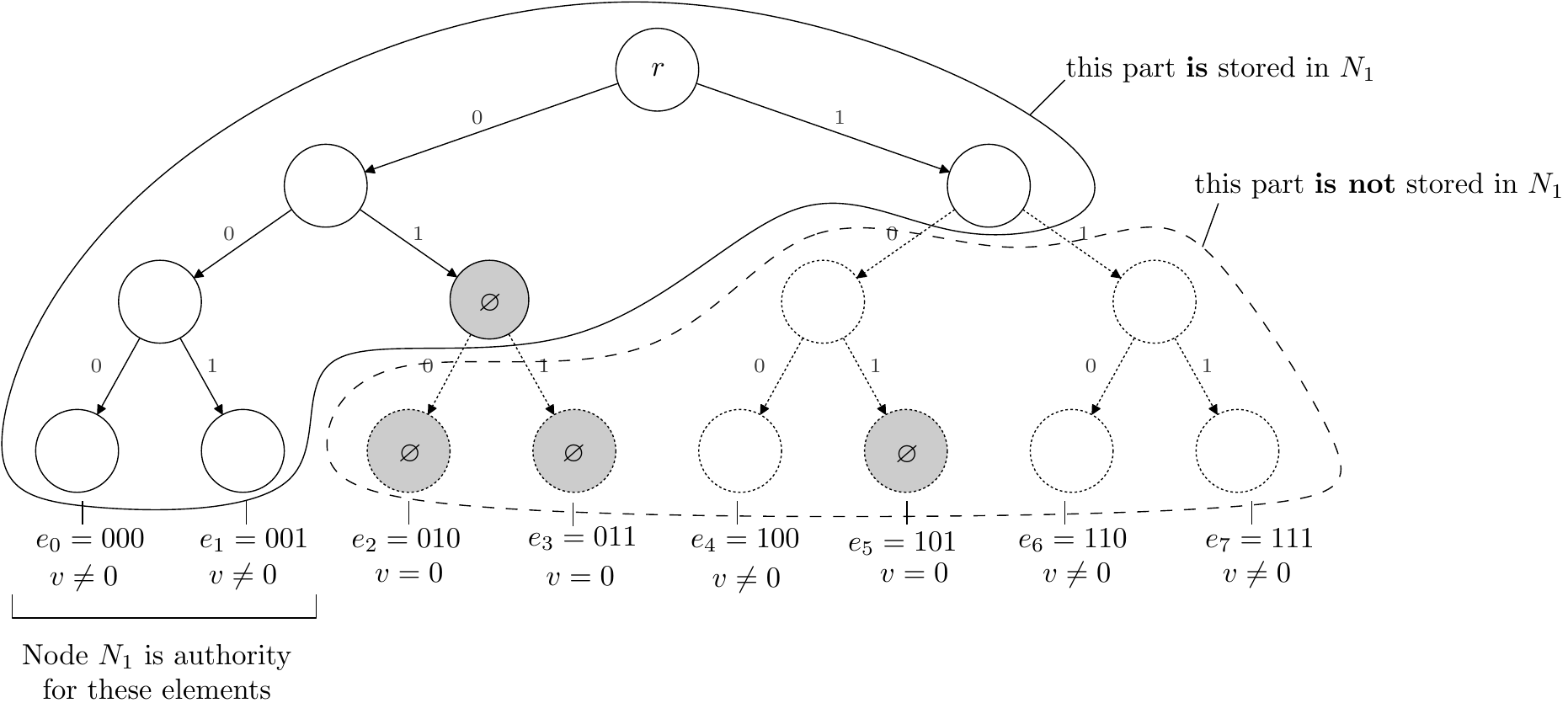}
		\subcaption{The pADS stored by $N_1$.}\label{fig:prefix-mht}
	\end{minipage}

	\begin{minipage}[b]{\linewidth}
		\centering
		\includegraphics[width=\linewidth]{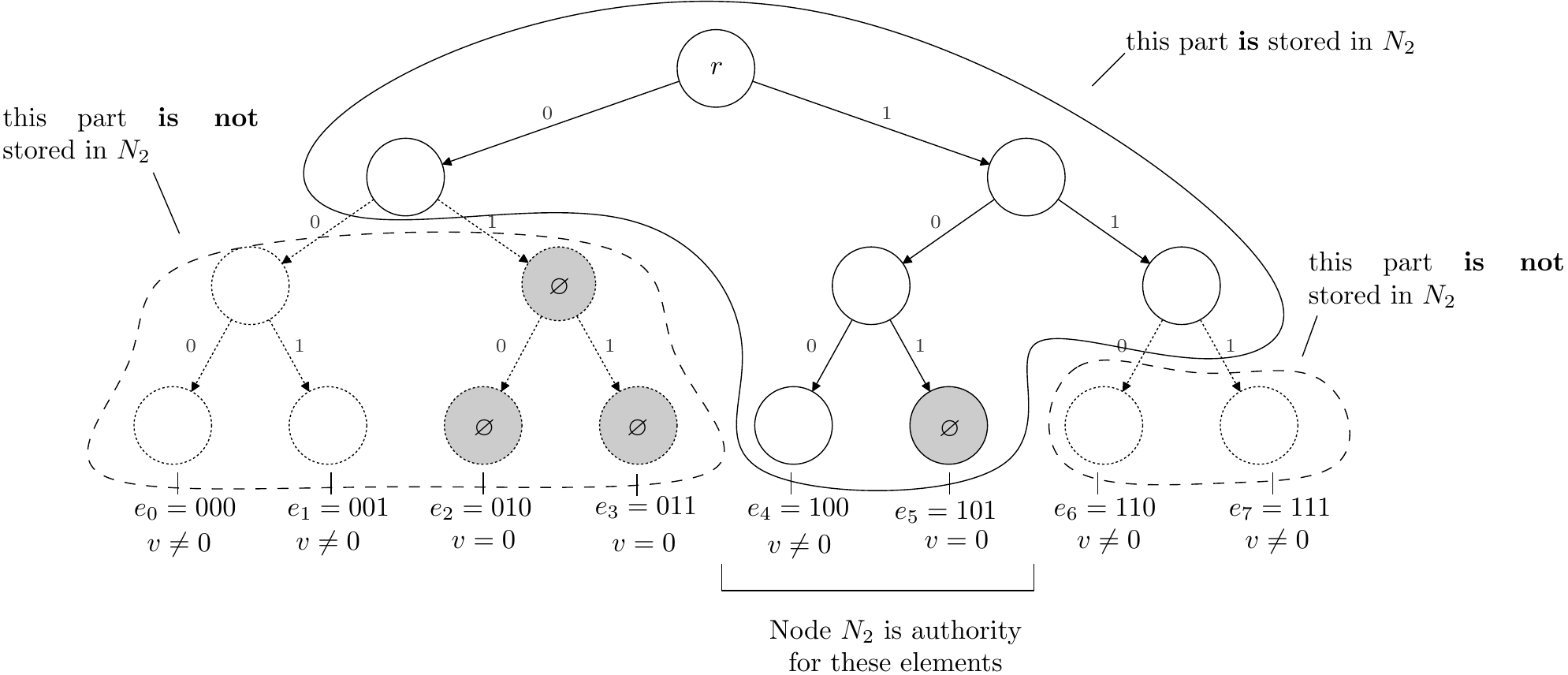}
		\subcaption{The pADS stored by $N_2$.}
		\label{fig:prefix-mht-second}
	\end{minipage}
	\caption{An example of pruned ADS (\emph{pADS}) for two nodes, each storing just a part of the same ADS.}
	\label{fig:two-prefix-mht}
\end{figure}

Since the storage is considered untrusted by the validation role, we equip the state with an
ADS. For simplicity, we assume the ADS to be constructed as a binary
prefix tree (see Figure~\ref{fig:two-prefix-mht}) where we suppose that
elements are identified by strings of bits of the same length.
Logically the state is represented as a complete binary tree where the leaves are all
possible state elements. As for any DHT, in general, not all the state
is kept by each node.  A node that stores the value of an element $e$
is called \emph{authority} for $e$. If node $N$ is not authority for $e$, we prune $e$ from the ADS stored by $ N $ using the technique
described in Section~\ref{sec:back.ADS}. We call this pruned version a \emph{pADS}.

Figure~\ref{fig:two-prefix-mht} depicts an example of an ADS, over state elements $ e_0, \dots, e_7 $, 
and how two nodes $N_1$ and $N_2$ store  different pruned versions of it.  All state elements are associated to a value.  
A special value, denoted $\varnothing$ in figure, refers to the hash associated with subtrees whose leaves store only
null values, like all unused elements. 
In the example, $ N_1 $ is authority for  $e_0$ and
$e_1$, while $N_2$ is authority for  $e_4$ and
$e_5$. The part of the ADS delimited by the solid line is the structure stored by a specific node, which we call pADS, while
the part delimited by the dashed line is pruned.

% Let a node $ N $, $ N_1 $ in figure, it is possible to detect two region in the ADS. The first region delimited by the solid line is the part of the ADS stored in $ N_1 $.

 %A special value (denoted $\varnothing$ in Figure~\ref{fig:prefix-mht})
%refers to the hash associated with subtrees whose leaves store only
%null values, like all unused elements. 

%The part of the pADS drawn with dotted lines is not stored by the node.

Each node $N$ stores a pADS for the blockchain state in a certain instant of time. 
In a synchronous scenario, all pADSes are the pruned version of the same ADS. In a real scenario, this is no longer true because each node receives the updates with a different delay. For this reason, each node $N$ stores a pADS for the blockchain state that is associated with a block $b_i$ 
(i.e., the state before the execution of the transactions in 
$b_i$). We call that block the \emph{pivot block} for $N$.

%Figure~\ref{fig:two-prefix-mht} shows two node ($ N_1 $ and $ N_2 $) that have the same pivot block, so the same ADS, but they are authority for different elements, therefore they store a different pADS.

% When new blocks are added to the blockchain, also the pivot block is advanced 
% by applying changes according to the execution of the transactions of the new 
% pivot $b_{i+1}$.\todo{$b_i$?}
When a new block is received and validated by $ N $, it is added to the local view of the blockchain. At the same time, the pADS is updated by applying 
changes according to the execution of the transactions of $b_{i}$, and 
$b_{i+1}$ becomes the new pivot, where $ b_{i+1} $ may be the last block in the local view of $N$ of the blockchain or an older block still stored by $ N $ (see Section~\ref{sec:validation}).
Since block propagation in the network takes some time, nodes can have a slightly skewed view of the
state. This aspect is taken into account within the validation role.  

A query for element $e$ can be answered by a node $N$ that is authority for 
$e$. It returns the value $v$ of $e$, the proof $p$ obtained by its pADS (see 
Section~\ref{sec:background}) and the index of the pivot block for $N$.

\subsection{Transaction Creation}

A transaction \emph{involves} a number of state elements.  These are
the state elements read and updated by the operations executed in the
transaction. Operations may be complex, like
in an invocation of a smart contract. However, we constrain the
operations to act only on a set of state elements that should be known
before the execution of the transaction.

In our approach, a transaction is similar to one in
traditional blockchains. It specifies sender(s), receiver(s),
operations, and all involved state elements. As in traditional
blockchains, for state elements that contain cryptocurrency balances
to be charged, a signature is also provided to prove that the sender
of the transaction is also the owner of charged state elements.

%\begin{figure}[h]
%	\centering
%	\includegraphics[width=\linewidth]{figures/transaction_creation_proof}
%	\caption{The pADS of the node $ N_a $, authority for the elements $ e_0 $ and $ e_1 $.
%		Let $ e_1 $ be the element involved in a transaction, the proof $ p $ is the collection of hashes of green node}
%	\label{fig:ledger}
%\end{figure}

Suppose a node $N_c$ intends to create a new transaction $t$. The set
of the state elements involved in $t$ is denoted by $E$. To know
the current value associated with each $ e \in E$, $N_c$ queries the
DHT obtaining, from a node $N_a$ authority for $e$, the tuple
$\delta_{i}(e)=\langle v, p, i\rangle$, where $v$ is the value for $e$
known to $N_a$, $p$ is the proof obtained from the pADS of $N_a$, and
$i$ is the index of the pivot block of $N_a$. When $N_c$ broadcasts
$t$, it attaches $\delta_{i}(e)$ to $t$ for each $e$. These additional
data are attached without any additional signature, since they will be
verified against root-hashes taken from blocks, and will be used
during mining by nodes executing the validation role.

\subsection{Validation Role of a Node}\label{sec:validation}

\begin{figure}[h]
	\centering
	\includegraphics[width=\linewidth]{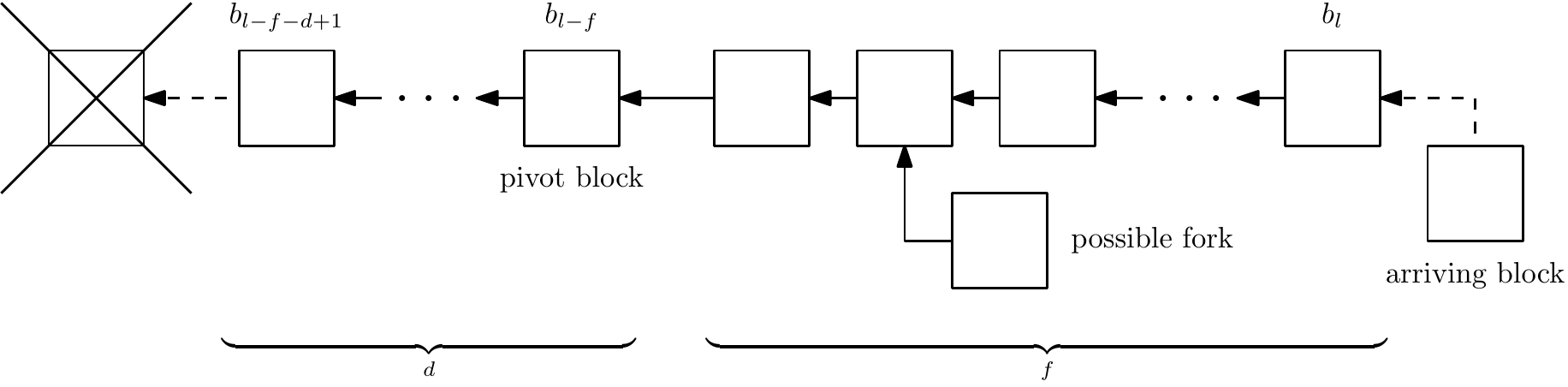}
	\caption{A node, for its validation role, stores only the last $d+f$ blocks 
	(see text). In the figure, $l$ denotes the index of the last block.}
	\label{fig:ledger}
\end{figure}

For the validation role, each node $N$ stores only the last $d + f$ blocks it 
received (see Figure~\ref{fig:ledger}). We denote $\varLambda_N$ this 
\emph{truncated blockchain} of $N$. At a certain instant of time, the truncated 
blockchain of distinct nodes may be different, since the propagation of new 
blocks in the network may take some time.  We suppose that each branch of a 
fork can be at most $f$ blocks long. Hence, the block at depth greater than $f$ 
cannot be undone by a fork resolution. The pivot block for the storage role for 
a certain node $N$ is chosen to be $b_{l-f}$, where $l$ is the index of last 
block in $\varLambda_N$.

We suppose that $d$ is big enough so that the proofs attached with
each transaction are computed on states related to blocks that are still
in $\varLambda_N$ when $N$ validates the transaction for the
mining of a new block. The dimensioning of $d$ should take into
account the time taken to create
a transaction, comprising the query to the DHT,  the propagation
delay of the transaction, and the maximum time a transaction has to wait 
to be included in a block in case of a peak of requests (see also Section~\ref{sec:discussion}).

\begin{figure}
	\centering
	\includegraphics[width=\linewidth]{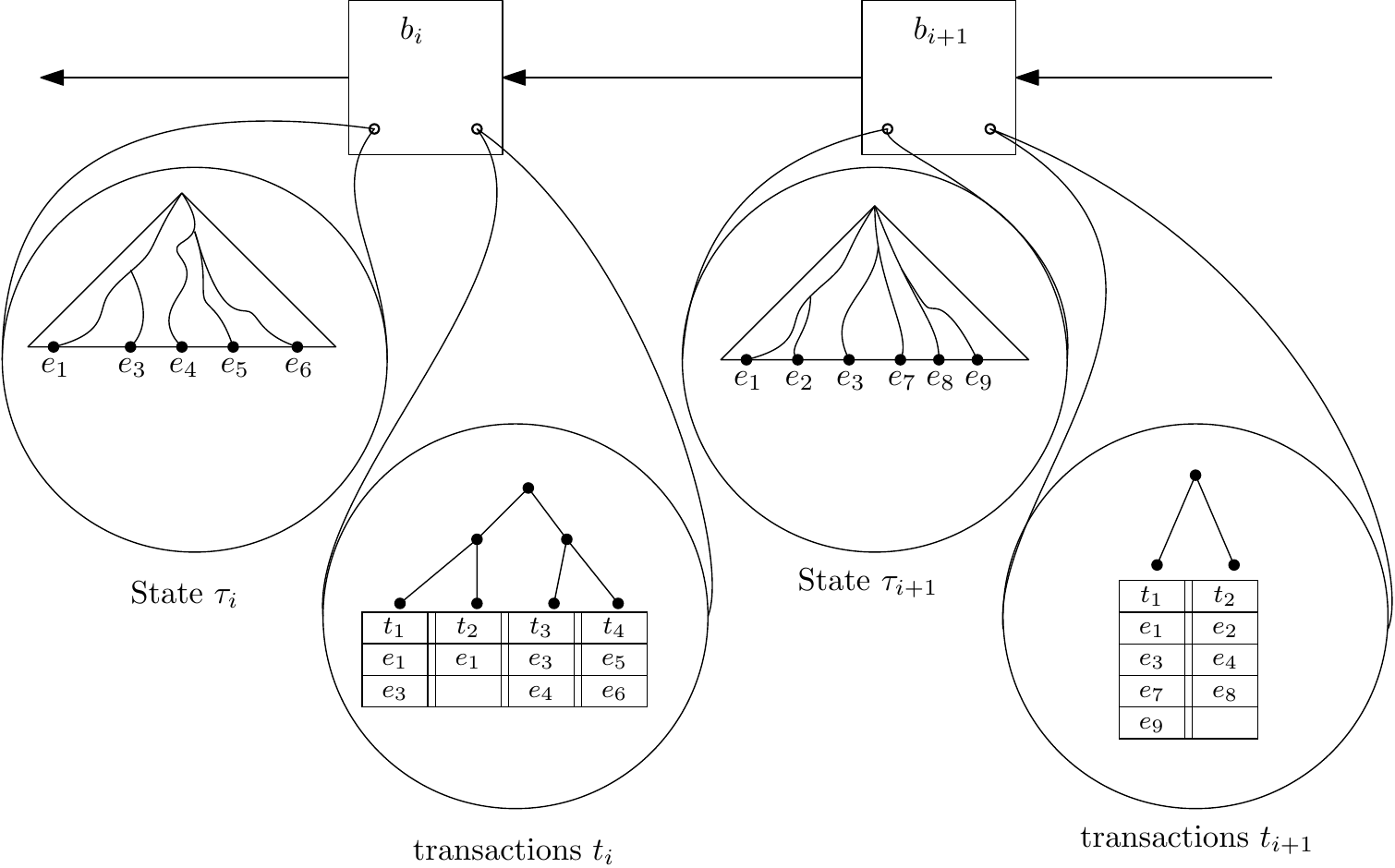}
	\caption{Content of two consecutive blocks. Each block $b_i$ contains a 
	pruned representation $\tau_i$ of the state before the execution of the 
	transactions in $b_i$. In $\tau_i$, the only unpruned leaves are the state elements 
	involved in the transactions of the block.}
	\label{fig:blockState}
\end{figure}

Each block $b_i$ contains a number of transactions. In traditional blockchains 
these are stored in a Merkle Tree whose root-hash is stored in the block. We 
depict it in this way in Figure~\ref{fig:blockState}, even though this is not 
relevant for our approach. The union of all involved state elements in
transactions associated with $b_i$ is denoted $E_i$. The block $b_i$ also 
contains a pADS $\tau_i$ representing the state of the blockchain before the 
application of the transactions in $b_i$, but only elements in $E_i$ are 
explicitly represented in $\tau_i$, the rest is pruned. Note that, the 
root-hash of $\tau_i$ is uniquely associated with the whole state at a certain time. 
We also denote by $ \pi_i $ a distinct pADS (with the same topology of
$\tau_i$ but different content) obtained applying all transaction in $
b_i $ to the state represented by $ \tau_i $. Note that, a node can
choose to store $ \pi_i $'s or re-compute them when needed.

\paragraph{Validation of a block} Let $b_l$ be the last block of 
$\varLambda_N$. When $ N $ receives a new block $ b_{l+1} $, it checks its 
validity considering the root-hash of $\pi_{l}$, which should turn out to be 
equal to the root-hash of $ \tau_{l+1} $. Each transaction in  $ b_{l+1} $ is 
then validated considering its execution starting from the values of elements 
as in $ \tau_{l+1} $, according to the consensus rules.  If all consensus rules 
are respected, $ b_{l+1} $ is appended to $\varLambda_N$, $b_{l-f-d}$ is 
removed from $\varLambda_N$ and $l$ is incremented.

Since the pivot block is now changed, the pADS for the storage role of $N$ is 
updated accordingly. Namely, $\tau_{l-f-1}$ is updated with the changes of the 
execution of transitions of $b_{l-f-1}$ into a $\bar \tau$. Then, $\bar \tau$ 
is used to update values of leaves or hashes of pruned subtrees in the pADS 
used for the storage role.

\paragraph{Creation of a block} To mine a new block, a node $N$ chooses a set of
\emph{selected transactions} to be included in the candidate next block
$b_{l+1}$, based on some criteria (e.g., prioritising higher fees). We
call $E_{l+1}$ the state elements that are involved in at least one of
the selected transactions. For each $e \in E_{l+1}$, all
$\delta_{i}(e)=\langle v, p, i\rangle$ attached to the selected
transactions are checked for their integrity by comparing the
root-hash computed from $p$ and $v$ against the root-hash of
$\tau_{i}$, which should be in $\varLambda_N$ if $d$ is big enough.

We now compute the pADS $\bar\tau = \tau_{l+1}$ to be included into
$b_{l+1}$ representing the state before the execution of the selected
transactions. Note that, $ \pi_l $ can not be used as $ \tau_{l+1} $ since they have a different pruning.
 We start by creating a $\bar{\tau}$ with the final
structure but with no hashes and values. This is done by merging all
the proofs attached to the selected transactions considering only
topological information and ignoring hashes and values. Then, we
iteratively perform the following operations starting from $ x = l$
and decrementing $ x $. We consider $\pi_x$ and, only for nodes
with unset hash/value in $\bar\tau$, set their hashes/values with the content of the
corresponding nodes in $\pi_x$. We then do the same thing with nodes of proofs in
$ \delta_{x}(e) $ that are attached to selected transactions, again
filling only empty hashes/values. We iterate until $ \bar{\tau} $ is
fully populated. At the end, $ \bar{\tau} $ turns out to be equal to $\tau_{l+1}$.
This procedure ends since, in the worst case, $\bar\tau$ is
populated only by all $ \delta_{i}(e) $ attached to selected transactions.
Its correctness derives from the fact it gives precedence to the most up-to-date
hashes/values and up-to-date hashes should take into account 
non updated hashes/values by the way transactions update the state.

Each selected transaction $t$ is then validated considering its execution starting 
from the values of elements as in $\tau_{l+1}$, according to the consensus 
rules. If all consensus rules are met, $t$ is actually added to the block. The new 
block $b_{l+1}$ contains all the valid transactions $t$ and 
$\tau_{l+1}$ (pruned from elements only used by invalid transactions). Then $ N $ executes the consensus 
algorithm and, if/when successful, broadcasts the block to the blockchain network.

	\section{Discussion}\label{sec:discussion}

% su https://medium.com/@albpalau/analyzing-the-hardware-requirements-to-be-an-ethereum-full-validated-node-ii-415b8457c6e8
% c'è una tabella di confronto con vari setup.
% In generale possiamo stimare tra i 4 e i 10 giorni per sincronizzare
% (usando Parity)
% 
% analizzando i dati di https://etherscan.io/charts e di https://bitinfocharts.com/ethereum/:
% - block count
%   = ~ 6.8 M
% - block size
%   = 9.7 kB (media aritmetica da etherscan)
%   = 18 kB (ultimo valore da bitinfocharts)
% - block time medio
%   = 15.6 s (secondo etherscan)
%   = 14.3 s (secondo bitinfocharts)
% - database size(?) secondo bitinfocharts
%   = 667.10 GB
% - blockchain di un client geth in modalità fast
%   = ~ 118 GB (aggiornato al 7 dicembre)
%     geth fast -> https://github.com/ethereum/go-ethereum/pull/1889
% - numero di indirizzi totali
%   = ~ 490 k
%
% dimensione di una transazione:
% - minima: 110 byte
%    questa ha 2 byte di dati extra -> https://etherscan.io/getRawTx?tx=0xc5eee3ae9cf10fbee05325e3a25c3b19489783612e36cb55b054c2cb4f82fc28
% - massima: 780 kB
%    secondo il limite imposto sul gas per blocco -> https://ethereum.stackexchange.com/questions/1106/is-there-a-limit-for-transaction-size
%    vedi -> https://etherscan.io/tx/0x25e54394ab4e5f17d6e1240c02c1a6c4bb675ef9471f1105b006988f5fe5aec1
% - average: ~ 190 byte
%

In this section, we discuss several aspects of a possible realisation
of the described approach. We assess our approach on the basis of
some parameters taken from the Ethereum network. Currently, in
Ethereum, a broadcasted block has an average size of 18 kB and 
a transaction has an average size of 200 bytes.
Supposing to adopt our approach, additional information should be attached 
to transactions and blocks. In both cases, its size depends on the number of 
involved state elements.

In our approach, each broadcasted transaction attaches values and
proofs of the involved state elements, relative to a certain pivot
block. Since values usually have negligible size, we focus on proofs.
For each state element, in our very simple construction of a pADS, a
proof has as many elements as the number of bits of the identifier of
the state element. For example, if an identifier has 160 bits and a
hash is 20 bytes, we obtain for each proof a size of 3200 bytes.
However, depending on the design, many state-elements that are
involved in a transaction may be close to each other and thus share most of
the path to the root. For example, in Ethereum the state elements
which represent the ``storage'' of a contract account are all in the same
subtree. A set of proofs can be represented itself as a pADS.
Further, the pADS may be a patricia trie compressing long chains. For
this reason, we assume an average of 500 bytes for each involved
state element.
In the Ethereum case, most transactions are calls to smart contracts, generally
involving more that two state elements. We assume 5 as an average
number of state elements per transaction. With these assumptions, 
each transaction turns out to be about 2700 bytes, which is one order of magnitude
larger than the standard one.
We leave a more precise estimation as a future work.

%Thus, each payment transaction, with 
%2 state elements, gets 
%an average overhead of about 6.5 kB. 
%
%, while supposing  \st{for payment transactions and 16.2 kB for smart 
%contract calls}.

% overhead pagamenti: (32 + 3200) * 2
% overhead contract:  (32 + 3200) * 5

%A payment 
%transaction involves 2 state elements, but, since in Ethereum most transactions 
%call smart contracts, we assume an average of 5 involved state elements, 
%leaving a more precise estimation as a future work. 

A broadcasted block $b_i$ contains $\tau_i$, which is a pADS covering
all state elements involved in transactions contained in $b_i$. Hashes
of inner nodes of $\tau_i$ are omitted since they can be computed from
their leaves (see Section~\ref{sec:back.ADS}). We estimate the size of $\tau_i$
as the size of the union of all the proofs of involved state elements in $b_{i}$.
In Ethereum, a block has 90
transactions on average. Given the assumptions above, we estimate an average of
450 involved state elements per block. Thus, each block gets an average
overhead of 225 kB, again one order of magnitude bigger than the standard block.

We now estimate the time needed by a node to synchronise. This is the
time needed to receive $d+f$ blocks. Usually, in Ethereum, a
transaction is considered confirmed after 3 minutes which is about 12
blocks, hence, we set $ f = 12 $. Regarding the dimensioning of $d$,
for simplicity, we consider a model in which propagation of broadcast
communications in the network takes some time but is synchronised, in
the sense that all nodes receive the same data at the same time. In
this model, all nodes have the same pivot block. A node $N_c$, to create a
transaction $\theta$, queries the DHT to obtain proofs of
state elements involved in $\theta$. Suppose that the first query is
served at time $t_0$ with index $i$. Then, all queries are completed
within time $t_0 + \Delta t_\mathrm{DHT}$. Supposing transactions
propagate in the network in time $\Delta t_\mathrm{pr}$, $\theta$
arrives to a miner $N_m$ at time $t_1 = t_0 + \Delta t_\mathrm{DHT} +
\Delta t_\mathrm{pr}$. For $N_m$ to be able to check the proofs
attached with $\theta$, $b_i$ should still be in $\varLambda_{N_m}$ at
$t_1$. This is verified if $\Delta t_\mathrm{DHT} + \Delta
t_\mathrm{pr} < (d-1) T$, since $\varLambda_{N_m}$ is updated every block
time $T$. The work described in~\cite{steiner2010evaluating} reports a
look-up time for Kademlia of no more than 30 seconds. To take into
account the time to broadcast a new transaction and the time it has to
wait in queue, we conservatively set $d=8$ (which is equivalent to 2
minutes). %In~\cite{decker2013information} is reported an
% exponentially decreasing distribution of the propagation delay of a
% block.
%After 40 seconds there are still $ 5\% $ of nodes which have not
% received the block yet. For this reason, we set $ d=8 $ (which is
% equivalent to 2 minutes).
Thus, the data to be downloaded to set up a node from scratch is about
4.8 MB, which at the speed of 2 Mb/s takes about 19 seconds. A very
high speed up with respect to the time currently taken by an Ethereum
node, which is in the order of hours at best.

This speed up is not totally for free. Our approach also increases the time
taken by the sender node to prepare the transaction, due to queries to
be performed on the DHT. We note that those queries, one for each
state element involved in the transaction, can be performed in parallel. 
Further, for use cases in which nodes should perform a large amount of similar transactions, specific caching mechanisms can be designed.

We note that, a transaction involving a large number of state elements is a considerable burden for the network, due to the fact that corresponding proofs should be stored in the block (i.e., in $ \tau_i $). Hence, a blockchain adopting our approach can discourage this kind of transactions by suitable means, like higher fees or gas consumption. On the other hand, a reward may be given to nodes that commit more 
storage for the DHT or that serve more DHT requests.

Concerning security, if we assume that blocks in $ \varLambda_N $ are trusted, the proofs that are received along with the transactions to be processed are enough to guarantee the authenticity of the involved state elements. Transactions containing proofs with indexes outside $ \varLambda_N $ are discarded. Validation of broadcasted blocks only requires the previous block to be trusted, which is a common assumption for blockchains. 

Since resynchronisation of a node is easy, we expect a high rate of disconnection and reconnection of nodes. During reconnection, a node needs a way to be sure that the first block it downloads belongs to the right blockchain, for example to the same blockchain that it was attached before reconnection. This problem is mitigated in regular blockchains by the fact that the nodes download the whole chain and have the hash of the first block hard-coded. In our approach, to avoid eclipse attacks during synchronisation, we may introduce \emph{checkpoint blocks}, whose hash is supposed to be kept by nodes for some time and included in all (or some) of the following blocks. This method allows for safe resynchronisation after a disconnection for a limited amount of time.

	\section{Conclusions}\label{sec:conclusions}

We have presented a method to significantly decrease the time taken by a node of a blockchain network to download the data needed to properly work.  In particular, validation can be performed storing a small amount of data.
To achieve this, we take advantage of authenticated data structures and flexibly distribute the storage of the blockchain state on a DHT made by the same nodes of the network. The main feature of this methodology is its speed, which allows the first synchronisation of a node to be performed in less than a minute.
Additionally, our approach is independent from the adopted consensus algorithm and from the consensus rules.

As future works, we plan to prototypically implement this approach in a software derived from one of the major blockchain networks (e.g., Ethereum). 
We plan to perform an extensive test and possibly proposing the idea to the corresponding community. We also consider interesting to investigate the relation of our approach with other scaling techniques, like sharding.

	\bibliographystyle{plain}
	\bibliography{references}

\begin{thebibliography}{10}

\bibitem{benet2014ipfs}
Juan Benet.
\newblock Ipfs-content addressed, versioned, p2p file system.
\newblock {\em arXiv preprint arXiv:1407.3561}, 2014.

\bibitem{chauhan2018blockchain}
Anamika Chauhan, Om~Prakash Malviya, Madhav Verma, and Tejinder~Singh Mor.
\newblock Blockchain and scalability.
\newblock In {\em 2018 IEEE International Conference on Software Quality,
  Reliability and Security Companion (QRS-C)}, pages 122--128. IEEE, 2018.

\bibitem{danzi2018analysis}
Pietro Danzi, Anders~Ellersgaard Kalor, Cedomir Stefanovic, and Petar Popovski.
\newblock Analysis of the communication traffic for blockchain synchronization
  of iot devices.
\newblock In {\em 2018 IEEE International Conference on Communications (ICC)},
  pages 1--7. IEEE, 2018.

\bibitem{kokoris2018omniledger}
Eleftherios Kokoris-Kogias, Philipp Jovanovic, Linus Gasser, Nicolas Gailly,
  Ewa Syta, and Bryan Ford.
\newblock Omniledger: A secure, scale-out, decentralized ledger via sharding.
\newblock In {\em 2018 IEEE Symposium on Security and Privacy (SP)}, pages
  583--598. IEEE, 2018.

\bibitem{martel2004general}
Charles Martel, Glen Nuckolls, Premkumar Devanbu, Michael Gertz, April Kwong,
  and Stuart~G Stubblebine.
\newblock A general model for authenticated data structures.
\newblock {\em Algorithmica}, 39(1):21--41, 2004.

\bibitem{maymounkov2002kademlia}
Petar Maymounkov and David Mazieres.
\newblock Kademlia: A peer-to-peer information system based on the xor metric.
\newblock In {\em International Workshop on Peer-to-Peer Systems}, pages
  53--65. Springer, 2002.

\bibitem{nakamoto2008bitcoin}
Satoshi Nakamoto.
\newblock Bitcoin: A peer-to-peer electronic cash system.
\newblock 2008.

\bibitem{steiner2010evaluating}
Moritz Steiner, Damiano Carra, and Ernst~W Biersack.
\newblock Evaluating and improving the content access in kad.
\newblock {\em Peer-to-peer networking and applications}, 3(2):115--128, 2010.

\bibitem{tamassia2003authenticated}
Roberto Tamassia.
\newblock Authenticated data structures.
\newblock In {\em European Symposium on Algorithms}, pages 2--5. Springer,
  2003.

\bibitem{tamassia2005efficient}
Roberto Tamassia and Nikos Triandopoulos.
\newblock Efficient content authentication over distributed hash tables.
\newblock In {\em Proc. Int'l Conf. Applied Cryptography and Network Security
  (ACNS'07),}. Citeseer, 2005.

\bibitem{wood2014ethereum}
Gavin Wood.
\newblock Ethereum: A secure decentralised generalised transaction ledger.
\newblock {\em Ethereum project yellow paper}, 151:1--32, 2014.

\end{thebibliography}

\end{document}